\documentclass[12pt]{article}
\usepackage{a4wide}
\usepackage{amsmath,amsthm,amsfonts}
\usepackage{graphicx}
\usepackage{bbm}
\usepackage{color}

\newcommand{\url}[1]{{\tt \small #1}}
\newcommand{\Ex}[2]{\mathbb{E}_{#1}\!\left[\,#2\,\right]}
\newcommand{\ExT}[3]{\mathbb{E}_{#1}^{#2}\!\left[\,#3\,\right]}

\newcommand{\one}{\ensuremath{\mathbbm{1}}}
\newcommand{\beq}[0]{\begin{equation}}
\newcommand{\eeq}[0]{\end{equation}}
\newcommand{\beqn}[0]{\begin{equation*}}
\newcommand{\eeqn}[0]{\end{equation*}}
\newcommand{\balign}[0]{\begin{aligned}}
\newcommand{\ealign}[0]{\end{aligned}}
\newcommand{\beqan}[0]{\beqn\balign}
\newcommand{\eeqan}[0]{\ealign\eeqn}
\newcommand{\beqa}[1]{\beq\label{#1}\balign}
\newcommand{\eeqa}[0]{\ealign\eeq}
\newcommand{\eref}[1]{(\ref{#1})}

\title{{\large \bf Parsimonious HJM Modelling\\ for Multiple Yield-Curve Dynamics}
}
\author{
Nicola Moreni\thanks{Banca IMI, {\tt nicola.moreni@bancaimi.com}}
\ \ \
Andrea Pallavicini\thanks{Banca Leonardo, {\tt andrea.pallavicini@bancaleonardo.com}}
}
\date{\small First Version: July 16, 2010.  This version: \today}

\begin{document}

\maketitle

\begin{abstract}
For a long time interest-rate models were built on a single yield curve used both for discounting and forwarding. However, the crisis that has affected financial markets in the last years led market players to revise this assumption and accommodate basis-swap spreads, whose remarkable widening  can no longer be neglected. In recent literature we find many proposals of multi-curve interest-rate models, whose calibration would typically require market quotes for all yield curves. At present this is not possible since most of the quotes are missing or extremely illiquid. 
Thanks to a suitable extension of the HJM framework, we propose a parsimonious model based on observed rates that deduces yield-curve dynamics from a single family of Markov processes. Furthermore, we detail a specification of the model reporting numerical examples of calibration to quoted market data.
\end{abstract}

{\bf JEL classification code: G13. \\ \indent AMS classification codes: 60J75, 91B70}

\medskip

{\bf Keywords:} Yield Curve Dynamics, Multi-Curve Framework, Gaussian Models, HJM Framework, Interest Rate Derivatives, Basis Swaps, Counterparty Risk, Liquidity Risk.

\newpage
{\small \tableofcontents}
\vfill
{\footnotesize \noindent The opinions here expressed  are solely those of the authors and do not represent in any way those of their employers.}
\newpage

\section{Introduction}
\label{sec:introduction}

Classical interest-rate models were formulated to satisfy by construction no-arbitrage relationships, which allow to hedge forward-rate agreements in terms of zero-coupon bonds. As a direct consequence, these models predict that forward rates of different tenors are related to each other by strong constraints. In practice, these no-arbitrage relationships might not hold. An example is provided by basis-swap spread quotes, which are  significantly non-zero, while they should be equal to zero if such constraints held.

This is what happened starting from summer 2007, with the raising of the credit crunch, where market quotes of forward rates and zero-coupon bonds began to violate the usual no-arbitrage relationships in a macroscopic way, under both the pressure of a liquidity crisis, which reduced the credit lines, and the possibility of a systemic break-down suggesting that counterparty risk could not be considered negligible any more. The resulting picture, as suggested by Henrard (2007), describes a money market where each forward rate seems to act as a different underlying asset.

\begin{figure}[t]
\begin{center}
\includegraphics[width=0.9\textwidth]{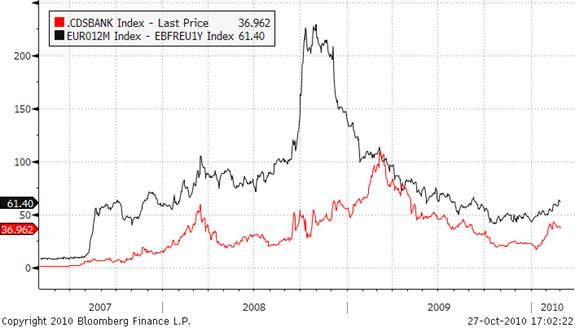}
\end{center}
\caption{
Historical series of Euribor-1y minus Eurepo-1y spread (black line) and a synthetic index formed by senior one-year CDS of a basket of twelve representative European banks (red line) ranging from May 2007 up to March 2010. Values are in basis points. The figures are obtained from Bloomberg\textsuperscript{\textregistered} platform.
\label{fig:cds}
}
\end{figure}

There are empirical studies supporting the idea that Libor rate levels cannot be utterly justified by counterparty credit risk arguments. In a European Central Bank working paper, Eisenschmidt and Tapking~(2009) compare the spread of the Euribor over the general collateral repo rate to the spread of banking-sector credit default swaps of the same tenor during the crisis period. Authors found that there is evidence of a large, persistent and time varying component of the Euribor-Eurepo spread that cannot not be explained by counterparty credit risk. In figure \ref{fig:cds} we show the historical series of Euribor-Eurepo spread for a rate tenor of one year and of a synthetic index composed by senior one-year CDS spread of a basket of twelve European banks representative of the Libor panel. Surely the two series have some common qualitative characteristics. Yet, we find that the sharp rise in the Euribor-Eurepo spread of September 2008 is only found three-four months later in the CDS spread series, confirming that a liquidity crisis needs time to evolve as credit crisis. Hence, counterparty risk is only  one of the Libor dynamics driving factors,  as discussed in Heider{\it et al.}~(2009).

Recently in the literature some authors started to deal with these issues, mainly concerning the valuation of cross currency swaps as Boenkost and Schmidt (2005), Kijima et al. (2009). In these papers, as in Henrard (2007, 2009), the problem is faced in a pragmatic way by considering each forward rate as a single asset without investigating the microscopical dynamics implied by liquidity and credit risks.  Attempts in this different direction are made in Morini (2009), Morini and Prampolini (2010) and Fries (2010). In particular, we refer to Moreni and Morini (2010) where Libor rates of different tenors are microscopically associated to different short rates, which, in turn, are obtained by adding an instantaneous credit spread to the risk-free short rate.

Besides microscopic approaches, many authors extended yield-curve bootstrapping to a multi-curve setting, eventually resulting in new pricing models. These latters are often inspired by other asset classes, as Bianchetti (2009), Chibane and Sheldon (2009), Kijima {\it et al.}~(2009), Mercurio (2009), Mart{\`\i}nez (2009), Kenyon (2010), or Pallavicini and Tarenghi (2010). We cite also a slightly different approach by Fujii {\it et al.}~(2010) and Mercurio (2010), where each basis spread is modelled as a different process.

However, the hypothesis of introducing different underlying assets may lead to over-parametrization issues that affect  the calibration procedure. Indeed, the presence of swap and basis-swap  quotes on many different yield curves is not sufficient, as the market quotes swaption premia only on few yield curves.  For instance, even if the Euro market quotes one-, three-, six- and twelve-month swap contracts, liquidly traded swaptions are only those indexed to the three-month (maturity one-year) and the six-month (maturities from two to thirty years) Euribor rates. Swaptions referring to other Euribor tenors or to Eonia are not actively quoted. A similar line of reasoning holds also for caps/floors and other interest-rate options.

In this paper we wish to introduce a parsimonious model which is able to describe a multi-curve setting by starting from a limited number of (Markov) processes. Among the classical single yield-curve models, this goal is achieved by the HJM framework by Heath, Jarrow and Morton~(1992), and by the functional Markov models by Hunt, Kennedy and Pelsser~(2000), where a single family of Markov processes is used to drive all the interest-rate derived quantities. Our proposal is to extend the logic of the former (HJM) to describe with a family of Markov processes all the curves we are interested in.

%
%

The structure of the paper is the following: Section~\ref{sec:ratesfundations} reviews the fundamental money-market concepts that underlie 
the construction of a multi-curve framework; in Section~\ref{sec:extending} we describe an original extended HJM framework able to handle 
many yield curves; in Section~\ref{sec:calibration}, we detail a simple yet relevant specification (dubbed the Weighted Gaussian Model) of the model that allows for simple evaluation of plain vanilla options, together with the results of its calibration to market data; finally, Section~\ref{sec:conclusions} reviews our contributions and hints for further developments.

\section{Multi-curve relevant features}
\label{sec:ratesfundations}
In order to motivate our modelling choices, it is useful to summarize the changes that 
occurred because of the credit crunch and the crucial issues a multi-curve framework should face. In this section we start identifying the risk-neutral measure, i.e. the risk-free discount term-structure, with the one coming from Overnight Indexed Swaps (OIS), and then we introduce risky rates (Libor). We also discuss, supporting our arguments with empirical analysis, the monotonicity properties of basis spreads and multi-tenor Libors. 

\subsection{Risk-free rates}
First of all we assume that the market is arbitrage free, hence postulating the existence of a risk-neutral measure. Under this measure every (risk-free) tradable asset instantaneously increases its value at the risk-free  rate $r_t$. Furthermore, we introduce (risk-free) zero-coupon bond prices and instantaneous  forward rates as
\begin{equation}\label{eq:1curveDef}\begin{aligned}
P_t(T)& := \Ex{t}{-\int_t^Tdu\,r_u}
\\
f_t(T)& := \ExT{t}{T}{r_T}
\end{aligned}\end{equation}
where the first expectation is taken under risk-neutral measure, and the last expectation is taken under a measure whose numeraire is $P_t(T)$ (hereafter simply $T$-forward measure).\\

As usual, we wish to link our risk-free rates to market quotes. 
In classical single-curve interest-rate models, zero-coupon bond prices observed at time $t=0$ form a term structure
\[
T \mapsto P_0(T)
\]
which can be made consistent with a selection of quotes (deposits, futures and interest-rate swaps).
However, since the beginning of the crisis, many of them have been carrying a relevant amount of credit and/or liquidity risk and cannot be considered as belonging to the risk-neutral economy. Thus, the subset of  the instruments to bootstrap the risk-free term structure from has to be carefully chosen. A closer look at the Euro money market makes clear that quoted instruments are indexed on three reference indices\footnote{See European Banking Federation site at {\tt http://www.euribor-ebf.eu}$\,$.}: Eonia, Euribor and Eurepo.
\begin{itemize}
\item Eonia is an effective rate calculated from the weighted average of all overnight unsecured lending transactions undertaken in the interbank market.
\item Euribor(s) are offered rates at which Euro interbank term deposits of different maturities are traded by one prime bank to another one.
\item Eurepo(s) are offered rates at which Euro interbank secured money market transactions are traded.
\end{itemize}

Eonia and Euribor rates are unsecured, so that they incorporate the default risk of the counterparty of the transaction, while Eurepo rates are secured and free of credit risk. Thus, Eurepo rates could seem  the natural proxy for risk-free rates\footnote{See for instance Eisenshmidt and Tapking (2009) where the Euribor-Eurepo spread is used as an indicator of credit risk.}. The main issue with Eurepo is that the longest quoted instrument  has a maturity of one year. Longer maturities Euro money market deals are only indexed on Euribor and Eonia indices. In particular, we found Eonia swap contracts (OIS) up to thirty years. Because of the plurality of available OIS instruments and of the reduced credit/liquidity exposure on overnight deposits,  to many extent Eonia rates are the best available proxy for risk-free rates. This point has been stressed by many authors, and we refer  to Fujii {\it et al.}~(2010) for more detailed arguments. 

\subsection{Libor rates}
\label{sec:libratesnlimit}
It is a common habit to refer to unsecured deposit rates over the period $[t,T]$  as Libor rates ($L_t(T)$). In this paper we follow this nomenclature and we reserve the term Euribor for the index used as reference rate for deposits in the Euro area. 
As usual we can introduce  the forward rates $F_t(T,x)$ defined as
\begin{equation}\label{eq:fwdLibDef}
F_t(T,x) := \ExT{t}{T}{L_{T-x}(T)}.
\end{equation}
Forward rates $F_t(T,x)$ are by construction martingales under the $T$-forward measure and  each of them represents the par rate seen at $t$ for a swaplet 
accruing over $[T-x,T]$ and paying at $T$ a fixed rate in exchange for $L_{T-x}(T)$.

Notice that accordingly to what said in the previous section,  we consider one-day deposits as being risk-free, while 
the longer the tenor, the greater will be the credit charge on unsecured deposit rates. In other words
we are thinking Eonia rates as (non-quoted) one-day-tenor Libor rates reducing as much as possible the deposit risks .
By pushing this analogy further we interpret  Libor rates as microscopic rates at the same level of the short-rate, and write 
\begin{equation}\label{eq:limitLr}
r_t = \lim_{x\rightarrow 0} L_t(t+x),
\end{equation}
which, given Eqs. (\ref{eq:1curveDef}) and (\ref{eq:fwdLibDef}), also reads
\begin{equation}\label{eq:limitFf}
f_t(T) = \lim_{x\rightarrow 0} F_t(T,x).
\end{equation}

\subsubsection{Credit risk premium and liquidity issues}

The usual no-arbitrage relationship between (risk-free) zero-coupon bond prices and Libor rates holds only for non-defaultable counterparties and instruments without liquidity risk. Hence, if $L_t(x)$ is a Libor rate related to the period $[t,t+x]$, we get in general
\[
L_t(t+x) \not= \frac{1}{x}\left(\frac{1}{P_t(t+x)}-1\right)
\;,\quad
\forall x > 0\; .
\]
Hence, when the presence of credit and liquidity risks invalidate the possibility of replicating Libor indexed deposits with
 non-risky bonds $P_t(T),$ then interest-rate modelling should consider Libor rates of different tenors as different assets. Yet, they should not move apart in a random way. At first glance, credit risk arguments imply that deposits with longer tenor must be charged for a higher risk premium, so that, if the risk-free yield curve is non decreasing, forward-rates should be a non-decreasing function of $x.$

\begin{figure}[t]
\begin{center}
\includegraphics[width=0.9\textwidth]{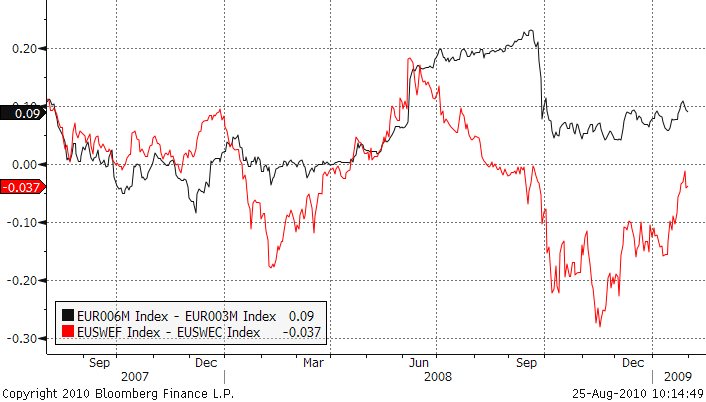}
\end{center}
\caption{
Historical series of Euribor-6m minus Euribor-3m spread (black line) OIS-6m minus OIS-3m spread (red line) ranging from August 2007 up to February 2009. The figures are obtained from Bloomberg\textsuperscript{\textregistered} platform.
\label{fig:steep}
}
\end{figure}

For instance, let us consider the EUR money market and focus on the risk-free yield curve bootstrapped from Eonia indexed products, such as OIS up to one year of maturity. We identify risk-free linearly compounding rates with single-period  OIS rates defined as  
$$
E_t(T,x):=\frac{1}{x}\left(\exp\left\{\int_{T-x}^T\!\!du\,f_t(u)\right\}-1\right)\,.
$$ 
If the risk-free curve is not inverted, credit risk premium arguments should lead to
\[
E_0(x,x) > E_0(x',x') \Longrightarrow L_0(x) > L_0(x')
\;,\quad
\forall x>x'.
\]

However, liquidity issues may invalidate such relationships. Actually  let us recall that both Eonia and Euribor rates refer to unsecured contracts, but  Euribor rates do not represent effective transactions, while Eonia rate does. As an example of violations, we plot in figure \ref{fig:steep} the daily historical values of spreads $s_E:=E_0(6m)-E_0(3m)$ and $s_L=L_0(6m)-L_0(3m).$
We notice that in periods of great turmoil, as the last trimester of 2007, even if the risk-free yield curve was often non-inverted, 
still the $s_L$ happened to be negative.\\

As a consequence, in the following we will  not impose direct constraints on Libor or forward rates, 
focusing on relationships to link forward-rate volatilities.

\subsubsection{Basis-swap spreads}

\begin{figure}[t]
\begin{center}
\includegraphics[width=0.9\textwidth]{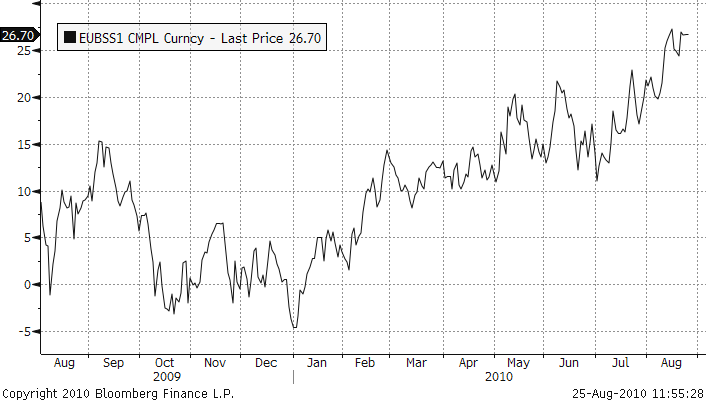}
\end{center}
\caption{
Historical series of basis-swap spread between Euribor-12m and Euribor-6m ranging from August 2009 up to August 2010. The figures are obtained from Bloomberg\textsuperscript{\textregistered} platform.
\label{fig:basis}
}
\end{figure}

The starting point of our analysis was the raise of basis-swap spreads after the credit crisis. Once again simple  credit risk arguments would require basis-swap spreads to be positive, but liquidity issues should also be considered. In the Euro area basis-swaps are quoted with maturities ranging from one year up to thirty years, so that each leg contains a strip of many Euribor rates. The averaging effect weakens the liquidity impact, but does not cancel it out. Indeed, in figure \ref{fig:basis} we see that the basis-swap spread between one year and six-month Euribor rates was often negative in the last trimester of 2009.

However, we find that in most cases quotes of basis-swap spreads are positive. In the literature only Fujii {\it et al.}~(2010) and Mercurio~(2010) force constraints on basis-swap spreads by direct modelling basis-swap spreads with respect to the EONIA rates with non-negative processes, but this condition does not ensure that all quoted basis-swap spreads remain positive.\\

Within our modeling framework, it will be quite difficult to impose constraints on basis-swap spreads positivity, since we will
not model them directly.

\section{Extending the HJM framework}
\label{sec:extending}
Our goal is to extend the classical HJM framework to include curves associated to different tenors by modelling forward Libor rates by means of a common family of (Markov) processes. In the literature other authors proposed generalizations of the HJM framework, see for instance Chiarella (2010) or Carmona (2004). In particular, in recent papers Mart{\`\i}nez (2009) and Fujii (2010) extended the HJM framework to incorporate multiple-yield curves and to deal with foreign currencies. 

Our approach differs from the previous ones mainly on two relevant points. First, we model only observed rates as in Libor market model approaches, avoiding the introduction of quantities such as \emph{forecasting curve bonds or instantaneous rates}. Second, we consider a common family of processes for all the yield curves of a given currency, so that we are able to build parsimonious yet flexible models.

As a consequence of the discussion of previous sections, and in order to keep the model as simple as possible, let us summarize the basic requirements the model must fulfill:
\begin{itemize}
\item[i)] existence of a risk free curve, with instantaneous forward rates $f_t(T)$
\item[ii)]existence of Libor rates, typical underlying of traded derivatives, with associated forwards $F_t(T,x)$
\item[iii)]no arbitrage dynamics of the $f_t(T)$ and the $F_t(T,x)$ (both being $T$-forward measure martingales) ensuring the limit case of Eq.(\ref{eq:limitFf})
\item[iv)] possibility of writing both the $f_t(T)$ and the $F_t(T,x)$ as function of a common family of Markov processes.
\end{itemize}
While the first two requisites are related to the set of financial quantities we are about to model, the last two are conditions we impose on their dynamics, and will be granted by a befitting choice of model volatilities. 

\subsection{Generalized dynamics}

According to requirements i) and ii) we model risk-free forward instantaneous rates $f_t(T)$ and (risky) forward Libor rates $F_t(T,x),$ for which we choose, under the $T-$forward measure, the following SDE\footnote{See appendix \ref{sec:notation} for vector and matrix notation.}.
\begin{equation}\label{eq:mainSDE}\begin{aligned}	
df_t(T)& = \sigma^*_t(T) \cdot dW_t\\
\frac{d(k(T,x)+F_t(T,x))}{(k(T,x)+F_t(T,x))}&= \Sigma^*_t(T,x)\cdot dW_t\\
\sigma_t(T)&:= \sigma_t(T;T,0)\\
\Sigma_t(T,x)&:=\int_{T-x}^T\!\!\!\!du\, \sigma_t(u;T,x)\,,
\end{aligned}\end{equation}
where we introduced the family of  volatility (row) vector processes $\sigma_t(u;T,x),$  the (row) vector of independent Brownian motions $W_t,$ 
and the set of shifts $k(T,x)$ that are required to satisfy\footnote{This assumption may be generalized asking that $k(T,x)\approx \phi(T,x)/x$ for a function $\phi$ such that $\lim_{x\rightarrow 0}\phi(T,x)=1.$} 
$$
\lim_{x\rightarrow 0} \,x\, k(T,x) = 1,
$$
that is, $k(T,x)\approx 1/x$ if $x\approx 0.$ \\
The identification of the volatility of risk free instantaneous forward rates $f_t(T)$ with $\sigma_t(T;T,0)$ is easily justified if we explicitly integrate the SDE for $F_t(T,x)$
\begin{multline*}
F_t(T,x) = k(T,x)\left(\left(1+\frac{F_0(T,x)}{k(T,x)}\right)\exp\left\{-\frac{1}{2}\int_0^t\!\!\left||\Sigma_s(T,x)\right||^2ds+\int_0^t\!\!\Sigma^*_s(T,x)\cdot dW_s\right\}-1\right)
\end{multline*}
and take the limit $x\rightarrow 0,$ such that $\Sigma_s(T,x)\approx x\sigma_s(T;T,0) + \mathcal{O}(x^2)$ and the exponential may be expanded in series of $x.$

The particular choice of a shifted forward Libor dynamics ensures the limit of Eq.(\ref{eq:limitFf})
 and is formally equivalent to the evolution of risk-free simple rates $E_t(T,x),$ which are for instance shifted lognormal when standard HJM volatilities leads to  an Hull and White model. In literature, direct modelling of shifted forward rates is also considered in Eberlein and Kluge (2007) (see also references therein), and in Papapantaleon (2010).

By means of the change of numeraire technique we have that 
\[
dW_t^{(T)} = dW_t^{(rn)} - d\left\langle W^{(rn)}, \log P(T)\right\rangle_t =  dW_t^{(rn)} + \left(\int_t^Tdu\sigma_t(u;u,0)\right) dt
\]
where $W^{(T)}$ and $W^{(rn)}$ are standard Brownian motions under $T-$forward and risk-neutral measure, respectively.  
It is then straightforward to write the dynamics of forward Libor rates and instantaneous risk-free rates under the risk neutral measure as
\begin{equation}\label{eq:generalLnF}\begin{aligned}	
\frac{d(k(T,x)+F_t(T,x))}{(k(T,x)+F_t(T,x))}&= \Sigma^*_t(T,x)\cdot\left[\int_t^T\!\!du\,\sigma_t(u;u,0)dt +dW_t\right],\\
df_t(T) &= \sigma^*_t(T) \cdot\left[\int_t^T\!\!du\,\sigma_t(u;u,0)dt+ dW_t\right]
\end{aligned}\end{equation}
$W_t$ being a risk-neutral measure multidimensional standard Brownian motion. 

\subsection{Constraints on the volatility process}

Let us analyse more in detail the dynamics of the shifted forward Libors under risk-neutral measure.
By integrating the SDE over the time period $[0,t]$ we get
\begin{multline*}
\ln\left(\frac{k(T,x)+F_t(T,x)}{k(T,x)+F_0(T,x)}\right) = \int_0^t\!\!\Sigma^*_s(T,x)\cdot \left [dW_s - \frac{1}{2}\Sigma_s(T,x)ds
+ \int_s^T\!\!du\sigma^*_s(u;u,0)ds\right]\,.
\end{multline*}	

To ensure the tractability and a Markovian specification of the model, we extend the single-curve HJM approach of Ritchken and Sankarasubramanian (1995), by setting
\beqa{eq:separableVol}
&\sigma_t(u;T,x) := h_t\cdot q(u;T,x) g(t,u)\\
&g(t,u) := \exp\left\{-\int_t^u \!\!dy\,\lambda(y)\right\}\\
&q(u;u,0)=1\,,
\eeqa
where $h$ is a matrix adapted process, $q$ is a diagonal matrix deterministic function (i.e. $q_{ij}=q_i\one_{i=j})$ and $\lambda$ is a deterministic array function. The condition on $q$ when $T=u$ is needed to ensure that in the  limit case $x\rightarrow 0$ we recover the standard Ritchen-Sankarasubramanian separability condition. \\
By plugging the expression for the volatility into Eq.\eref{eq:generalLnF}, it is possible to work out
the expression ending up with the representation 
\begin{multline}\label{eq:separableLnF}
\ln\left(\frac{k(T,x)+F_t(T,x)}{k(T,x)+F_0(T,x)}\right) = \\G^*(t,T-x,T;T,x)\cdot \left(X_t+Y_t\cdot \left(G_0(t,t,T)-\frac{1}{2}G(t,T-x,T;T,x)\right)\right),
\end{multline}
where we have defined the It\^o stochastic process $X_t$ 
$$
X^i_t := \sum_{k=1}^N \int_0^t g_i(s,t)\left(h^*_{ik,s} dW_{k,s} + (h^*_sh_s)_{ik}\int_s^t\!\!\!dy\,g_k(s,y)ds\right)\, ,i=1,\ldots,N
$$
and the auxiliary matrix process $Y_t$
$$
Y^{ik}_t:= \int_0^tds \,g_i(s,t)(h^*_sh_s)_{ik}g_k(s,t)\quad i,k=1,\ldots,N
$$
with $X_0^i=0$ and $Y_0^{ik}=0$, as well as the vectorial deterministic functions
\beqan
&G_0(t,T_0,T_1):=\int_{T_0}^{T_1}\!\!\!dy\,g(t,y)  \\
&G(t,T_0,T_1;T,x):= \int_{T_0}^{T_1}\!\!\!dy\,q(y;T,x)g(t,y)\,.
\eeqan
The limit case $x\rightarrow 0,$ as previously detailed for general $\sigma_t(u;T,x)$ still holds and we may check that 
$f_t(T)=\lim_{x\rightarrow 0}F_t(T,x).$\\ 

\subsubsection{Dynamics of state variables}

Equation \eref{eq:separableLnF} is the analogous of standard HJM reconstruction formula and is the main result of our paper. 
Let us notice that it returns a reconstruction formula for forward Libor rates, while standard HJM one is based on bonds. 
This important feature is consistent with the requirement of a model capable to directly describe market relevant quantities.
  
Thanks to our assumption we are fully able to describe instantaneous forward rates (i.e. discounting curve bonds) and forward Libor rates 
once we know the state variables $\left\{X_t,Y_t\right\},$ which satisfy, under the risk neutral measure, the following coupled (S)DE
\beqan
&dX^i_t = 	\sum_{k=1}^N\left(Y^{ik}_t-\lambda_i(t)X^i_t\right)dt + h^*_t\cdot dW_t\\
&dY^{ik}_t = \left[(h^*_th_t)_{ik} - (\lambda_i(t)+\lambda_k(t))Y^{ik}_t\right]dt.
\eeqan

Let us notice that forward Libor diffusion pre-factors\footnote{Actually, starting from \eref{eq:separableLnF}, and switching to the terminal $Q^T$ measure, we have $$dF_t(T,x) = [\kappa(T,x)+F_t(T,x)]G^*(t,T-x,T;T,x)\cdot h^*_t\cdot dW_t\,.$$} $G(t,T-x,T;T,x)$ depend on the $q(u;T,x).$ This flexibility is a desirable feature, as it allows for a locally tuned dynamics for forward Libor rates, as we show in the next section.

\subsubsection{Exact calibration and sensitivities}

Our approach focuses on market quantities and leaves us the freedom of choosing the $q(u;T,x)$ and the $\kappa(T,x)$ such as to exactly calibrate a selection of market data. These free parameters are independent from the skew/smile patterns that endogenously come with the risk free HJM dynamics of Eonia single-period rates. This is a relevant advantage over other microscopic multi-curve models where the dynamics of microscopic quantities uniquely determines implied volatility patterns for rates of any tenor and maturity.

Thus, as relevant tenors and maturities form a discrete set (for instance $x\in\{1,3,6,12\}$ months), we may reasonably set $\kappa(T,x)$ to be a piece-wise-constant deterministic function to be exactly calibrated to a subset of caplet or swaption skews. Further, we have the same possibility for a subset of at-the-money caplet or swaption volatilities by properly defining the $q(u;T,x)$ process.

For instance, it is possible to set
\[
q(u;T,x):=\hat q(T,x)p(u) \,,
\]
with $\hat q$ a scalar function and $p$ an array function. With this prescription $\hat q$ may be used to exactly calibrate a subset of at-the-money quotes, while the array $p$ allows to select the subset of the $X$ that is relevant for the diffusion of $F_t(T,x)$. In this way we may associate the dynamics of a chosen rate to a selection of relevant ``diffusion modes''.

The possibility of an exact calibration to a subset of at-the-money caplet or swaption volatilities and skews easily allows for sensitivity computation, and is similar to what happens in stochastic local volatility models, see for instance Torrealba (2010), where, after having calibrated the parameters of the volatility process, the local term allows for an exact calibration to some relevant market quotes.

\subsection{Eonia simple rates}\label{sss:eoniasimplerates}
For sake of completeness we may also compute Eonia simple rates $E_t(T,x)$ by plugging the separable volatility form within 
the relationship 
$$
1+xE_t(T,x) = \exp\left\{ \int_{T-x}^T \!\! dy \, f_t(y)\right\}
$$ 
such that\begin{multline}\label{eq:separableLnE}
\ln\left(\frac{1+xE_t(T,x)}{1+xE_0(T,x)}\right) = \\G_0^*(t,T-x,T)\cdot \left(X_t+Y_t\cdot \left(G_0(t,t,T)-\frac{1}{2}G_0(t,T-x,T)\right)\right).
\end{multline}
 
Let us notice that if we set $q(u;T,x)\equiv 1,$ then $G_0(t,T-x,T)\equiv G(t,T-x,T,x,T)$ such that 
$F_t(T,x)$ and $E_t(T,x)$ would differ only in their shifts and initial values. If we moreover choose $\kappa(T,x)=1/x,$ 
we would obtain a model with perfect instantaneous correlation between Libors and Eonia simple rates in which 
$$
\frac{1+xF_t(T,x)}{1+xE_t(T,x)} = \frac{1+xF_0(T,x)}{1+xE_0(T,x)}\,,
$$ 
hence showing that the static correction  model of Henrard (2009) is a particular case of this extended HJM framework.

\subsection{Swap rates}\label{ss:swprates}
Our framework also allows us to derive an (approximated) expression for swap rates dynamics.\\
Let us consider a swap with a $x$ tenor floating leg and a $\bar x$ tenor fixed one paying at times $\{T_{a+1},\ldots,T_{b}\}$
and $\{T_{\bar a+1},\ldots,T_{\bar b}\},$ respectively. The swap par rate equating the two legs is
$$
S_t^{ab}(x,\bar x):=\frac{\sum_{k=a+1}^b\tau_kP_t(T_k)F_t(T_k,x)}{\sum_{k=\bar a+1}^{\bar b}\bar \tau_kP_t(T_k)}
$$
where the quantities with a bar refer to the fix leg. We introduce the weights $w$ as
\[
w_k^{ab}(t)(x,\bar x) := \frac{\tau_kP_t(T_k)}{\sum_{k=\bar a+1}^{\bar b}\bar\tau_kP_t(T_k)}
\]
and  perform the usual freezing (see Errais and Mercurio (2005)) technique to obtain, under the swap measure $Q^{ab}$, 
\beqan\balign
dS^{ab}_t(x,\bar x) &\approx \sum_{k=a+1}^b w_k^{ab}(t) dF_t(T_k,x)  \\
&= \sum_{k=a+1}^b w_k^{ab}(t)\left[\kappa(T_k,x)+F_t(T_k,x)\right]\Sigma^*_t(T_k,x)\cdot dW_t \\
&\approx \left(S_t^{ab}(x,\bar x) + \psi^{ab}(x,\bar x)\right)\sum_{k=a+1}^b \delta_k^{ab}\Sigma^*_t(T_k,x)\cdot dW_t\,,
\ealign\eeqan
where
\beqan\balign
\psi^{ab}(x,\bar x):=\frac{\sum_{k=a+1}^b\tau_kP_0(T_k)\kappa(T_k,x)}{\sum_{j=\bar a+1}^{\bar b}\bar\tau_jP_0(T_j)}
\ealign\eeqan

\[
\delta^{ab}_k(x) := \frac{\tau_kP_t(T_k)(\kappa(T_k,x)+F_0(T_k,x))}{\sum_{j=a+1}^b\tau_jP_t(T_j)(\kappa(x,T_j)+F_0(x,T_j))}\,.
\]

With similar arguments we get an expression also for basis swap spreads, since we have
\[
B^{ab}_t(x,x') =: S^{ab}_t(x,x') - S^{ab}_t(x',x')\,.
\]

\subsection{Volatility dynamics}

As in the single-curve HJM framework we can add a stochastic volatility process to our model by extending the filtration to include also the information generated by the volatility process. A popular choice is to model the matrix process $h_t$ by means of a square-root process (see for instance Trolle and Schwartz (2009) and reference therein).

We start by replacing the $h_t$ process by
\[
h_t := \sqrt{v_t} R^* 
\]
where $R$ is a lower triangular matrix, while the variance $v_t$ is a vector process whose dynamics under risk neutral measure is given by
\[
dv_t = \kappa \left( \theta - v_t \right)\,dt + \nu\sqrt{v_t} \,dZ_t
\;,\quad
v_0 = {\bar v}
\]
where $\kappa$,$\theta$,$\nu$,$\bar v$ are constant deterministic vectors, and $Z_t$ is a vector of independent Brownian motions correlated to the $W_t$ processes as given by
\[
\rho_{ii} \,dt := d\langle Z,W_i \rangle_t
\;,\quad
{\rm and~} \rho_{ij} = 0 \; {\rm for~} i \neq j
\]
where $\rho$ is a diagonal deterministic correlation matrix.

With this choice we get shifted Heston dynamics for market rates, so that we can calculate option pricing with usual Fourier transform techniques (see  Lewis (2001)).

\section{Model calibration and numerical examples}
\label{sec:calibration}

As shown in Pallavicini and Tarenghi (2010) there are evidences that the money market for Euro area has moved to a multi-curve setting for what concerns the pricing of plain-vanilla instruments like interest-rate swaps, but the situation is not so clear for derivative contracts, where the calibration of volatility and correlation parameters may hide the impact of which yield curve is used in pricing. In particular, this holds for CMS swaps and CMS options, while the swaption market has evidences of pricing in the old single-curve approach, although some contributors start quoting in multi-curve framework from September 2010.

On the other hand, the money market for Euro area does not quote options on all rate tenors. In Euro area only options on the six-months tenor are widely listed, while the three-months tenor is present only in few quotes (swaptions with one-year tenor and cap/floors with maturities up to two years), and options on the other rate tenors are missing. Thus, any model which requires a different dynamics for each term-structure, has the problem that market quotes cannot be found to fix all its degrees of freedom.

Here, we select a simple but realistic volatility specification for the multi-curve HJM framework, which we calibrate to at-the-money swaption prices quoted by ICAP\textsuperscript{\textregistered} on Bloom\-berg\textsuperscript{\textregistered} platform on 12 of August 2010. We limit ourselves to a very simple calibration data-set since we are interested only in highlighting a relevant property of our framework, namely the possibility to build all volatility term-structures starting from few market quotes. We address to Pallavicini and Tarenghi (2010) for further calibration details for a simpler model specification (which consider independently each yield curve).

In particular, we consider a simple extension of a two-factor Gaussian model (see G$2$++ model in Brigo and Mercurio (2006)), that we call Weighted Gaussian model (WG$2$++ model), since the terms depending on Libor tenors appear as multiplicative weights of the $X$ processes. Notice that we could use more than two factors (WG$n$++ model), or we could add stochastic volatility to calibrate also the swaption volatility smile, leading to a Weighted Heston model (WH$n$++ model).

\subsection{The Weighted Gaussian model}
As an example we introduce a simple specification of our generic HJM multi-curve approach, with different dynamics for each forward Libor rate.
In practice it is a generalization of a shifted n-factor Hull and White model associated to risk-free rates.\\
Let us  set the volatility process $h_t$ to be in the form
$$
h_t := \varepsilon(t) h R^*\,, 
$$
where $h$ is a diagonal constant matrix $h_{ij}=h_i\delta_{i=j},$ $R$ is a lower triangular matrix representing the pseudo-square root of a correlation matrix $\rho$, and 
we allow for a time varying common volatility shape in the form
\[
\varepsilon(t) := 1 + (\beta_0 - 1 + \beta_1t)e^{-\beta_2t}\,,
\]
where $\beta_0$, $\beta_1$, $\beta_2$ are three positive constants.
Microscopical Markov factors $X$ and $Y$ evolve under the risk free measure, as
\beqa{eq:wgn}
&dX^i_t = \left(\sum_{j=1}^n{Y^{ij}_t}-\lambda_i X^i_t\right)dt + \varepsilon(t)h_i d\hat W^i_t\\
&dY^{ij}_t = \left( \varepsilon^2(t)h_ih_j\rho_{ij} + (\lambda_i+\lambda_j)Y^{ij}_t\right)dt\\
&d\langle \hat W^i \hat W^j \rangle_t = \rho_{ij}dt 
\eeqa
where the $\lambda_i$ are non negative constants, and $d\hat W_t:=R^* \cdot dW_t$.\\
The risk free short rate is given as usual by
$$
r_t := f_0(t) +\sum_{k=1}^n X^i_t
$$
and the shift term $f_0(t)$ allows to recover $t=0$ risk free yield curve\footnote{As the $Y$ are deterministic, 
this model is often written by explicitly computing the $Y$-related quantities such as the drift of the $X.$ Those quantities are then 
incorporated into a generic shift.}. \\
As for the tenor-maturity factors $q$, we chose a maturity independent form of the type
$$
q_i(u;T,x):=e^{-x\eta_i}\,.
$$
Numerical tests are done with $n=2,$ hence leading to ten free parameters 
$$
\left\{\lambda_1,\lambda_2,h_1,h_2,\eta_1,\eta_2,\rho_{12},\beta_0,\beta_1,\beta_2\right\}\,,
$$
and for sake of simplicity we set $\kappa(T,x) = 1/x.$\\
By construction this model supports different forecasting curves and we bootstrapped
initial forward Libors $F_0(T,x)$ by means of the Eonia term structure (discounting) and different tenors Euribor term structures.
\subsection{Benchmark models}
In our numerical examples we compare the results of the WG$2$++ model with respect to other two HJM-like models, all with two driving factors and time-dependent volatilities. We discount flows by means of the Eonia term structure and use for forecasting purposes the Euribor term structures.\\
\begin{itemize}
\item The G$2$++ model of Brigo and Mercurio (2006). This is a single-curve (old-style) model which we extend to incorporate time-dependent volatilities via the common time-dependent factor $\varepsilon(t).$
It is obtained by setting $\eta_j\equiv 0, $ (i.e. $q\equiv 1,$)  and $F_0(T,x)\equiv E_0(T,x).$  For this model we use, as discounting and forwarding curve, a term structure obtained with old-style standard techniques from deposits, futures and swap rates.
\item The MMG model of Pallavicini and Tarenghi (2010). This is an uncertain parameter multi-curve model which we restrict to have only one scenario. It is obtained by setting $\eta_j\equiv 0,$ uses separate forwarding and discounting curves and, as discussed in Sect.\ref{sss:eoniasimplerates}, reduces to Henrard static correction model. It has eight free parameters ($\lambda_1$,$h_1$,$\lambda_2$,$h_2$,$\rho_{12}$,$\beta_0$,$\beta_1$,$\beta_2$) and 
uses the same curves as the Weighted Gaussian.
\end{itemize}

\begin{table}[t]
\begin{center}
{\small
\begin{tabular}{|c|c|cc|}\hline
           &      Eonia & \multicolumn{2}{|c|}{Euribor} \\
      Date &  zero rate &   3M rate &  6M rate \\\hline
 13-Aug-10 &   0.4770\% &            &            \\
 16-Aug-10 &   0.4343\% &   0.8990\% &   1.1540\% \\
 17-Aug-10 &   0.4174\% &   0.8990\% &   1.1536\% \\
 23-Aug-10 &   0.4456\% &   0.9002\% &   1.1519\% \\
 30-Aug-10 &   0.4379\% &   0.9036\% &   1.1507\% \\
 06-Sep-10 &   0.4323\% &   0.9087\% &   1.1502\% \\
 16-Sep-10 &   0.4472\% &   0.9180\% &   1.1500\% \\
 18-Oct-10 &   0.4743\% &   0.9460\% &   1.1530\% \\
 16-Nov-10 &   0.5046\% &   0.9500\% &   1.1540\% \\
 16-Dec-10 &   0.5275\% &   0.9550\% &   1.1560\% \\
 17-Jan-11 &   0.5464\% &   0.9540\% &   1.1560\% \\
 16-Feb-11 &   0.5624\% &   0.9710\% &   1.1625\% \\
 16-May-11 &   0.6031\% &   1.0176\% &   1.2094\% \\
 16-Aug-11 &   0.6357\% &   1.0829\% &   1.2827\% \\
 16-Aug-12 &   0.7806\% &   1.4838\% &   1.6889\% \\\hline 
\end{tabular}
~
\begin{tabular}{|c|c|cc|}\hline
           &      Eonia & \multicolumn{2}{|c|}{Euribor} \\
      Date &  zero rate &   3M rate &  6M rate \\\hline
 16-Aug-13 &   0.9866\% &   2.0188\% &   2.2145\% \\
 18-Aug-14 &   1.2261\% &   2.4872\% &   2.6658\% \\
 17-Aug-15 &   1.4595\% &   2.8650\% &   3.0248\% \\
 16-Aug-16 &   1.6768\% &   3.1778\% &   3.3186\% \\
 16-Aug-17 &   1.8705\% &   3.4018\% &   3.5233\% \\
 16-Aug-18 &   2.0402\% &   3.5240\% &   3.6264\% \\
 16-Aug-19 &   2.1851\% &   3.6114\% &   3.6957\% \\
 17-Aug-20 &   2.3158\% &   3.7665\% &   3.8339\% \\
 16-Aug-22 &   2.5474\% &   3.9389\% &   3.9812\% \\
 18-Aug-25 &   2.7750\% &   3.7835\% &   3.8062\% \\
 16-Aug-30 &   2.9349\% &   3.2752\% &   3.2921\% \\
 16-Aug-35 &   2.8907\% &   2.5129\% &   2.5297\% \\
 16-Aug-40 &   2.7639\% &   2.2390\% &   2.2559\% \\
 16-Aug-50 &   2.5716\% &   2.5083\% &   2.5252\% \\
 16-Aug-60 &   2.4562\% &   2.6407\% &   2.6576\% \\\hline
\end{tabular}
}
\end{center}
\caption{
Eonia term-structure expressed in term of ACT/360 zero-rates and Euribor term structures for three- six- month tenors expressed in terms of ACT/360 forward rates. Bootstrapping details can be found on Pallavicini and Tarenghi (2010). Data bootstrapped from market quotes observed on 12 of August 2010.
\label{tab:disc}
}
\end{table}

\subsection{Initial forwarding and discounting curves}

The initial yield curves can be bootstrapped from the money market quotes. We refer again to Pallavicini and Tarenghi (2010) and references therein for a complete discussion. Here, we adopt their methodology. In particular we use the Eonia term-structure to discount cash flows, as a proxy for the risk-free yield curve (see also Fujii et al. (2010) and Mercurio (2010)).

In table \ref{tab:disc} we show the Eonia term-structure expressed in term of ACT/360 zero-rates, and the Euribor term structures for three- six- month tenors expressed in terms of ACT/360 forward rates.

\begin{table}[t]
\begin{center}
\hspace*{-1.1cm}
{\small
\begin{tabular}{|c|cccccccccccc|}\hline
           &         1Y &         2Y &         3Y &         4Y &         5Y &         6Y &         7Y &         8Y &         9Y &        10Y &        15Y &        20Y \\\hline
        1Y &     0.27\% &     0.56\% &     0.85\% &     1.12\% &     1.37\% &     1.61\% &     1.85\% &     2.08\% &     2.30\% &     2.50\% &     3.44\% &     4.24\% \\
        2Y &     0.43\% &     0.83\% &     1.22\% &     1.59\% &     1.95\% &     2.30\% &     2.63\% &     2.95\% &     3.26\% &     3.56\% &     4.83\% &     5.95\% \\
        3Y &     0.51\% &     0.98\% &     1.44\% &     1.87\% &     2.31\% &     2.71\% &     3.11\% &     3.48\% &     3.84\% &     4.20\% &     5.61\% &     6.90\% \\
        4Y &     0.55\% &     1.07\% &     1.56\% &     2.02\% &     2.47\% &     2.92\% &     3.35\% &     3.77\% &     4.17\% &     4.55\% &     6.05\% &     7.42\% \\
        5Y &     0.58\% &     1.11\% &     1.63\% &     2.12\% &     2.59\% &     3.06\% &     3.51\% &     3.94\% &     4.37\% &     4.77\% &     6.37\% &     7.80\% \\
        6Y &     0.59\% &     1.15\% &     1.69\% &     2.20\% &     2.68\% &     3.15\% &     3.60\% &     4.04\% &     4.48\% &     4.89\% &     6.53\% &     8.02\% \\
        7Y &     0.60\% &     1.16\% &     1.72\% &     2.24\% &     2.74\% &     3.21\% &     3.67\% &     4.12\% &     4.57\% &     5.00\% &     6.69\% &     8.20\% \\
        8Y &     0.61\% &     1.18\% &     1.74\% &     2.26\% &     2.77\% &     3.25\% &     3.71\% &     4.17\% &     4.63\% &     5.07\% &     6.80\% &     8.33\% \\
        9Y &     0.62\% &     1.19\% &     1.75\% &     2.28\% &     2.79\% &     3.28\% &     3.75\% &     4.22\% &     4.68\% &     5.13\% &     6.88\% &     8.45\% \\
       10Y &     0.62\% &     1.19\% &     1.76\% &     2.29\% &     2.81\% &     3.31\% &     3.79\% &     4.24\% &     4.70\% &     5.14\% &     6.90\% &     8.49\% \\
       15Y &     0.59\% &     1.14\% &     1.69\% &     2.21\% &     2.70\% &     3.20\% &     3.68\% &     4.14\% &     4.59\% &     5.04\% &     6.72\% &     8.27\% \\
       20Y &     0.55\% &     1.07\% &     1.58\% &     2.08\% &     2.57\% &     3.05\% &     3.51\% &     3.95\% &     4.37\% &     4.75\% &     6.39\% &     7.81\% \\\hline
\end{tabular}
}
\end{center}
\caption{
At-the-money swaption prices quoted by ICAP\textsuperscript{\textregistered} on Bloomberg\textsuperscript{\textregistered} platform on 12 of August 2010. Underlying swap's tenor on columns, its starting time on rows. In Euro area swaptions with one-year tenor are claims to enter a swap whose floating leg is indexed with the three-month Euribor rate, while all the other refer to the six-month Euribor rate.
\label{tab:price}
}
\end{table}

\subsection{Swaption pricing formula}

Swaption prices are given by the following expectation value under risk-neutral measure
\[
\Pi^{ab}_t:=\Ex{t}{ \exp\left\{-\int_t^{T_a}du\,r_u\right\} A^{ab}(T_a;{\bar x}) (S^{ab}(T_a;x,{\bar x})-K)^+}
\]
Under the approximation of Section \ref{ss:swprates}, which is pretty good for ATM swaptions, we get a log-shifted dynamics for
swap rates such that 
\[
\Pi^{ab}_t=A^{ab}(t;{\bar x}) \,{\rm Bl}(K+\psi^{ab}(x,{\bar x}),S^{ab}(t;x,{\bar x})+\psi^{ab}(x,{\bar x}),\Gamma^{ab}(x))
\]
where the annuity $A^{ab}$ is given by
\[
A^{ab}(t;{\bar x}) :=  \sum_{k=a+1}^b {\bar \tau_k}P_t(T_k)
\]
and ${\rm Bl}(\cdot)$ is the usual Black formula with given strike, forward rate and volatility. In particular the volatility $\Gamma^{ab}$ is given by
\[
\Gamma^{ab}(x) := \sqrt{ (\Delta^{ab}(x))^* \,\Sigma^{ab}\, \Delta^{ab}(x) }
\]
with the deterministic vector $\Delta^{ab}$ defined as
\[
\Delta^{ab}(x) := h e^{-\eta x}\sum_{k=a+1}^b \delta^{ab}_k(x) \frac{e^{-\lambda T_{k-1}}-e^{-\lambda T_{k}}}{\lambda}
\]
and the deterministic matrix $\Sigma^{ab}$ defined as
\[
\Sigma^{ab} := \int_t^{T_a} du\, (e^{\lambda u})^* \,\rho\, e^{\lambda u} \varepsilon^2(u)\,.
\]

For our calibration examples we consider at-the-money swaption prices quoted by ICAP\textsuperscript{\textregistered} on Bloomberg\textsuperscript{\textregistered} platform on 12 of August 2010, as given in table \ref{tab:price}. Notice that in the Euro area swaptions with one-year tenor are claims to enter a swap whose floating leg is indexed with the three-month Euribor rate, while all the other refer to the six-month Euribor rate. Swaptions referring to other Euribor tenors or to Eonia are not actively quoted.

\begin{table}[t]
\begin{center}
\begin{minipage}{0.4\textwidth}
{\small
\begin{tabular}{|c|ccc|}\hline
           &    \multicolumn{3}{|c|}{Model} \\
   Pars    &    G$2$++ &        MMG &   WG$2$++ \\\hline
    $\lambda_1$ &     0.0008 &     0.0090 &     0.0073 \\
    $\eta_1$ &     -       &    -        &     0.1581 \\
    $h_1$ &     0.0132 &     0.0057 &     0.0059 \\\hline
    $\lambda_2$ &     0.0036 &     5.0077 &     4.7344 \\
    $\eta_2$ &     -      &     -      &     0.8894 \\
    $h_2$ &     0.0162 &     0.0318 &     0.0411 \\\hline
   $\rho_{12}$ &    -0.9488 &    -0.8396 &    -0.8577 \\\hline
$\beta_0$ &     1.0976 &     1.2811 &     1.3160 \\
$\beta_1$ &     1.5277 &     1.3109 &     1.3327 \\
$\beta_2$ &     0.4928 &     0.6059 &     0.5900 \\\hline\hline
 $\chi^2$ &      100\% &    22.08\% &    16.99\% \\\hline
\end{tabular}
}
\end{minipage}
\begin{minipage}{0.55\textwidth}
\includegraphics[scale=0.65]{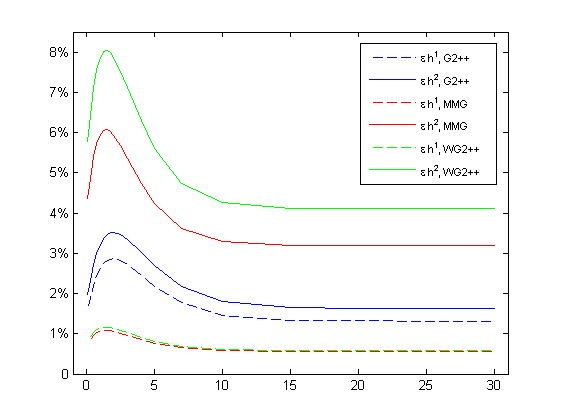}
\end{minipage}
\end{center}
\caption{
Model parameters obtained from the calibration procedure to at-the-money swaption prices quoted by ICAP\textsuperscript{\textregistered} on Bloomberg\textsuperscript{\textregistered} platform on 12 of August 2010. The Last row shows the calibration error normalized to the one obtained with the G$2$++ model. On the right panel the volatility backbone of each driving factor, namely the product $\varepsilon(t)h_k$ with $k\in\{1,2\}$ with respect to time $t$ in years.
\label{tab:calib}
}
\end{table}

\begin{figure}[t]
\begin{center}
\includegraphics[scale=0.6,trim=0 25 0 25,clip=true]{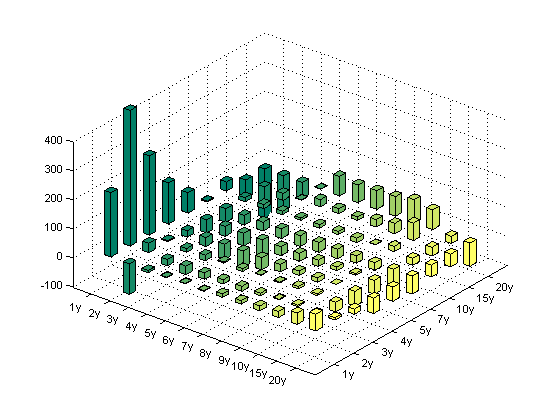}\\\hspace*{-1cm}
\includegraphics[scale=0.6,trim=0 25 0 25,clip=true]{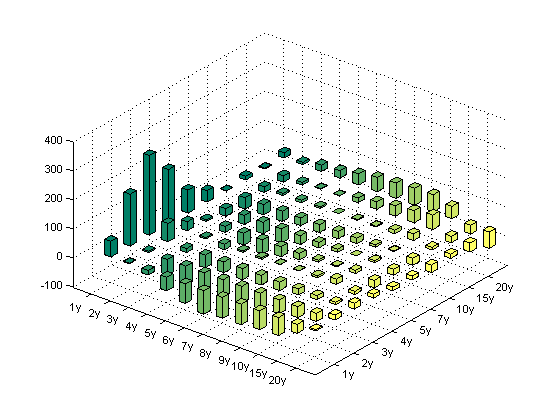}~
\includegraphics[scale=0.6,trim=0 25 0 25,clip=true]{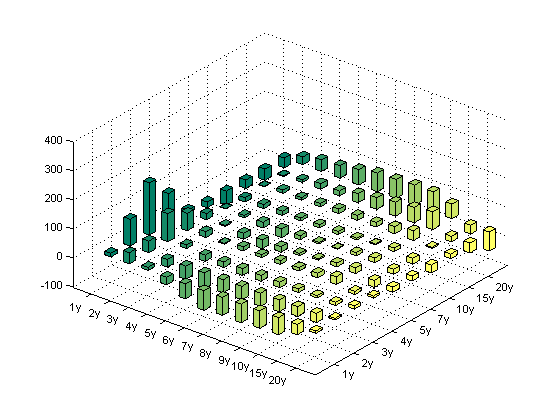}
\end{center}
\caption{
Differences in basis points between market and model-implied volatilities, namely calibration error in term of implied swaption volatilities. Upper panel is G$2$++ model, lower-left panel is MMG model, and lower-right panel is WG$2$++ model. Each panel shows on the left axis the underlying swap's tenor, while on the right axis its starting time.
\label{fig:calib}
}
\end{figure}

\subsection{Calibration examples}

In table \ref{tab:calib} we list the model parameters obtained from the calibration procedure. Notice that the two driving processes $X^1_t$ and $X^2_t$ operate on two different time scales. Indeed, by construction the first process has always a speed of mean reversion smaller than the one of the second process. This constraint is enforced while calibrating to avoid a degenerate problem.

In the figure on the right side of table \ref{tab:price} the volatility backbones of each driving factor, namely the product $\varepsilon(t)h_k$ with $k\in\{1,2\}$ plotted with respect to time $t$ in years. We can see that, allowing for more degrees of freedom along the Euribor tenor space as we increase the complexity of the model, the volatilities of the two driving processes $X^1_t$ and $X^2_t$ split apart: a higher volatility for the process with higher speed of mean reversion (process acting on a shorter time scale).

Calibration errors in term of implied swaption volatilities are shown in figure \ref{fig:calib}. We can see that the calibration error for swaptions with a tenor of one year is less as long as the model allows for incorporating multiple yield curves (MMG) and differentiating their dynamics (WG$2$++). Indeed, as previously stated in the Euro area swaptions with one-year tenor are claims to enter a swap whose floating leg is indexed with the three-month Euribor rate, while all the other refer to the six-month Euribor rate.

\begin{figure}[t]
\begin{center}
\includegraphics[scale=0.5]{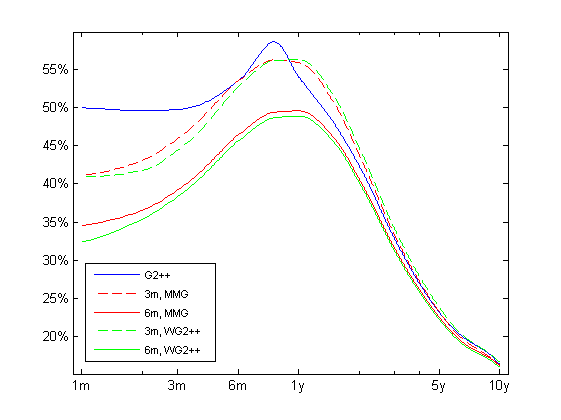}
\includegraphics[scale=0.5]{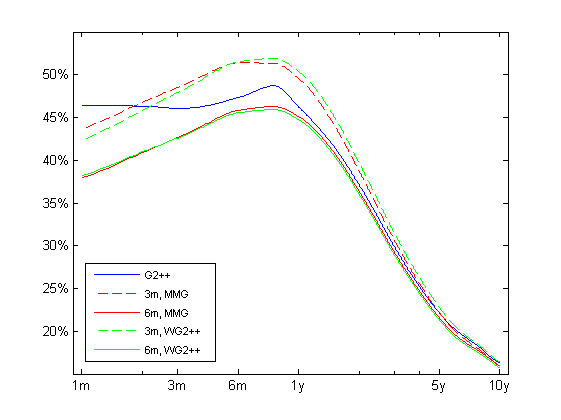}\\
\includegraphics[scale=0.5]{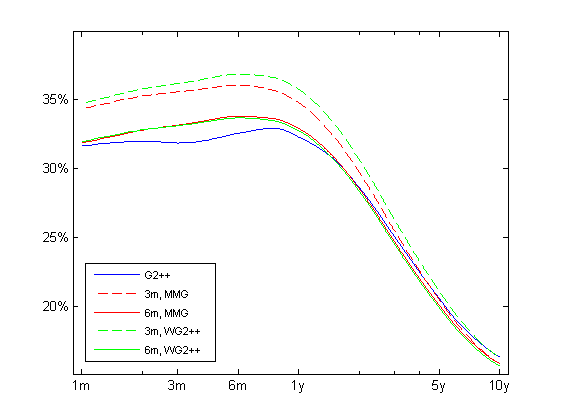}
\includegraphics[scale=0.5]{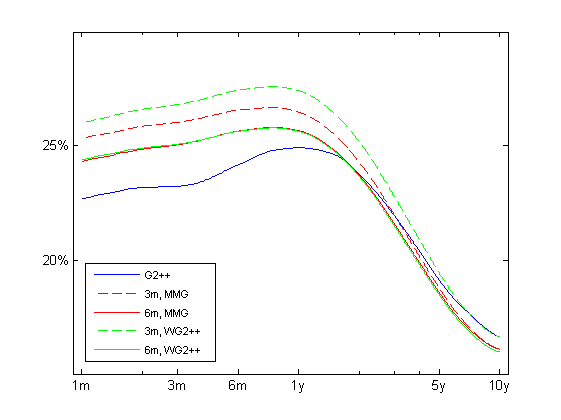}
\end{center}
\caption{
Each panel shows the implied volatilities by changing the underlying swap's starting time. Top-left panel is one-year underlying swap's tenor, top-right two-year tenor, bottom-left five-year tenor, bottom-right ten-year tenor.
\label{fig:vols}
}
\end{figure}

We show in figure \ref{fig:vols} the implied volatility for swaptions of different tenors and expiries as predicted by the WG$2$++ model and by the benchmark models. We show both claims to enter a swap whose floating leg is indexed with the three-month Euribor rate and the ones referring to the six-month Euribor rate. Notice that only the WG$2$++ model and the MMG model are able to differentiate between the two types of swaptions. In particular, we observe that only the WG$2$++ model is able to preserve such difference also for longer swaption tenors. Indeed, the MMG model produces the split because of different initial yield curves, while the WG$2$++ model relies also on a dynamical mechanism.

\section{Conclusions and further developments}
\label{sec:conclusions}

Interest-rate modelling requires a framework able to incorporate many initial yield curves, one for each Libor rate tenor plus one for discounting. Classical models may be extended in many ways, but, unfortunately, the market is too young to quote options on all tenors: only limited number of quotes are available and they are concentrated only in few tenors. Thus, a model, which allows a minimal extension of classical frameworks and, at the same time, allows for more complex dynamics when quotes will be available, is a relevant tool for both quants and practitioners.

In this paper we presented an extension of the HJM model which is able to deduce the dynamics of the discounting yield curve and of  market Libor rates of any tenor starting from a single family of Markov processes. Further, we calibrate a simplified version of the model, the Weighted Gaussian model, with two driving factors and deterministic volatility, to at-the-money swaption prices to size the effect of the new degree of freedom introduced to model different tenors. We also made a comparison with two benchmark models, already published in the literature, which turn out to be special cases of our model.

Our next step will be the calibration of a stochastic volatility version of the model, the Weighted Heston model, to the whole cube of swaption prices, and at the same time we hope the market evolves by quoting options on more tenors.

\appendix

\section{Appendix: vector and matrix notation}
\label{sec:notation}
When we consider a vector quantity $v$, we think it as a matrix with only one row, if a ``column" vector is needed we use the transposition operator, namely $v^*$. Further, we introduce also the vector whose entries are all of ones and we name it $\one$.

Let us consider two matrix quantities $a$ and $b$, whose elements are respectively $a_{ij}$ and $b_{ij}$ with $1\le i\le n$ and $1\le j\le m$. We define element-wise multiplication as the matrix $ab$ with elements:
\[
(ab)_{ij} := a_{ij}b_{ij}
\]
and, in the same fashion, also multiplication by a vector $v$, whose elements are $v_i$ with $1\le i\le n$, or a scalar $\kappa$ as 
\[
(v a)_{ij} := v_i a_{ij}
\;,\quad
(\kappa a)_{ij} := \kappa a_{ij}
\]
while index contraction as the matrix $a^*\cdot b$ with elements:
\[
(a^*\cdot b)_{jk} := \sum_{i=1}^n a_{ij}b_{ik}
\]

\end{document}